\documentclass[prl,superscriptaddress,twocolumn]{revtex4-1}
\usepackage{ulem,graphicx,amsmath,amssymb,epsfig,epstopdf}
\usepackage[colorlinks,linkcolor=blue,anchorcolor=blue,citecolor=blue,bookmarksnumbered]{hyperref}
\usepackage{mathrsfs}
\usepackage{amsbsy}
\usepackage{dcolumn}
\usepackage{color}
\usepackage{bm}
\def\be{\begin{equation}}
\def\ee{\end{equation}}
\def\bea{\begin{eqnarray}}
\def\eea{\end{eqnarray}}
\def\bes{\begin{subequations}}
\def\ees{\end{subequations}}
\newcommand{\ii}{\mathrm{i}}
\newcommand{\PT}{{\cal PT}}
\newcommand{\p}{{\cal P}}

\newcommand{\T}{{\cal T}}

\newcommand{\K}{{\cal K}}

\newcommand{\bA}{{\bm A}}
\newcommand{\bE}{{\bm E}}
\newcommand{\ba}{{\bm a}}

\newcommand{\bfr}{\mathbf{r}}
\newcommand{\rR}{{\rm R}}
\newcommand{\tb}{\tilde{b}}
\newcommand{\tba}{\tilde{\bm a}}

\newcommand{\bal}{{\bm \alpha}}
\newcommand{\tbal}{\tilde{\bm \alpha}}

\begin{document}
\title{Nonlinear Topological Edge States in a non-Hermitian Array of Optical Waveguides Embedded in an Atomic Gas}
\author{Chao Hang}
\affiliation{ State Key Laboratory of Precision Spectroscopy, East China Normal University, Shanghai 200062, China\\
NYU-ECNU Institute of Physics at NYU-Shanghai, Shanghai 200062, China}
\author{Dmitry A. Zezyulin}
\affiliation{ITMO University, St.~Petersburg 197101, Russia}
\author{Guoxiang Huang}
\affiliation{ State Key Laboratory of Precision Spectroscopy,
East China Normal University, Shanghai 200062, China\\
NYU-ECNU Institute of Physics at NYU-Shanghai, Shanghai 200062, China}
\author{Vladimir V. Konotop}
\affiliation{Departamento de F\'isica, Faculdade de Ci\^encias, Universidade de Lisboa, Campo Grande, Edif\'cio C8, Lisboa 1749-016, Portugal\\
Centro de F\'isica Te\'orica e Computacional,  Faculdade de Ci\^encias, Universidade de Lisboa, Campo Grande, Edif\'cio C8, Lisboa 1749-016, Portugal}
\date{\today}

\begin{abstract}
We propose a scheme comprising an array of anisotropic optical waveguides, embedded in a gas of cold atoms, which can be tuned from a Hermitian to an odd-$\PT$-symmetric configuration through the manipulation of control and assistant laser fields. We show that the system can be controlled by tuning intra- and inter-cell coupling coefficients, enabling the creation of topologically distinct phases and linear topological edge states. The waveguide array, characterized by a quadrimer primitive cell, allows for implementing transitions between Hermitian and odd-$\PT$-symmetric configurations, broken and unbroken $\PT$-symmetric phases, topologically trivial and nontrivial phases, as well as transitions between linear and nonlinear regimes. The introduced   scheme generalizes the Rice-Mele Hamiltonian for a nonlinear non-Hermitian quadrimer array featuring odd-$\PT$ symmetry and makes accessible unique phenomena and functionalities that emerge from the interplay of non-Hermiticity, topology, and nonlinearity. We also show that in the presence of nonlinearity the system sustains nonlinear topological edge states bifurcating from the linear topological edge states and the modes without linear limit. Each nonlinear mode represents a doublet of odd-$\PT$-conjugate states. In the broken $\PT$ phase,  the nonlinear edge states may be effectively stabilized when an additional absorption is introduced into the system.
\end{abstract}


\maketitle

{\it Introduction.--} Optical systems are universal simulators of physical phenomena from many areas of physics. In the past decade a particular attention was focused on optical analogues of topological insulators, having fundamental importance for condensed matter physics~\cite{elect1,elect2}. Pioneered by the works~\cite{Haldane2008a,Haldane2008b}, a new field of topological photonics~\cite{topphot1,topphot2} has emerged.
Non-Hermitian topological insulators were established to have two different types of phase transitions, i.e., the transitions between pure real and complex spectra of linear Hamiltonians~\cite{Bender1998,Bender2007} and between topologically distinct
phases~\cite{topphot1,topphot2}. Topological properties of linear non-Hermitian systems are now well understood, and their classifications based on the symmetries of systems are available~\cite{Gong18,Kawabata19}. It is also known that in a finite non-Hermitian system with boundaries, edge states can be sustained by nontrivial topological phases~\cite{topphot1,topphot2}.
Linear edge states at the interface between $\PT$-symmetric Su-Schrieffer-Heeger (SSH)~\cite{SSH} photonic lattices in distinct topological phases have been observed experimentally~\cite{Szameit}. Topological zero-energy edge states in passive-$\PT$ silicon waveguide arrays~\cite{Song19} have been reported, too.

In optical settings, topological phenomena are further enriched by nonlinearity~\cite{topphot3}.  Non-topological nonlinear parity($\p$) -- time ($\T$) symmetric systems~\cite{KYZ,Suchkov}
may sustain families of nonlinear modes without linear counterparts~\cite{ZezKonPRL2012}. Meantime, not any linear mode persists in the presence of nonlinearity, i.e., the nonlinearity must obey the symmetry consistent with that of the linear system in order to  enable a bifurcation of a nonlinear family from the linear limit~\cite{ZezKonPRL2012,ZezKon13}. Nonlinearity may also result in $\PT$-symmetry breaking~\cite{nonlin_PT,Lumer}, in pitchfork symmetry-breaking bifurcations \cite{Yang2014}, and in destabilizing (stabilizing) a linear mode which is otherwise stable (unstable)~\cite{ZezKonPRA2012}. Self-induced topological transitions and edge states have been reported in nonlinear SSH arrays~\cite{HKA2016}. 
Nonlinear topological edge modes were created in an array of pumped resonators~\cite{Kivshar}. 

However, so far nonlinear modes in non-Hermitian systems have been considered mainly under even $\PT$-symmetry, for which $\T^2=1$~\cite{KYZ,Suchkov}. It was shown that the realization of {\it odd} time reversal (for which $\T^2=-1$) is available~\cite{Konotop2018,Konotop2019}
by using the polarization of light in waveguides with anti-$\PT$-symmetric~\cite{Ge2013,Wu2015} coupling. 
The major difference between the even- and odd-$\PT$-symmetric lattices consists in their elementary cells: a primitive cell of an even-$\PT$-symmetric chain is a dimer~\cite{dimer,KYZ,Suchkov} whereas that of an odd-$\PT$-symmetric lattice is a quadrimer \cite{Konotop2018}
(in analogy with structures of  wave-functions of even- and odd-$\PT$-symmetric quantum Hamiltonians~\cite{Jones-Smith2010,Bender2011}). Particularly, guided modes in odd-$\PT$-couplers waveguides feature intrinsic symmetry-protected degeneracy which allows to manipulate superpositions of degenerate modes and results in unconventional   bifurcations of nonlinear states \cite{Konotop2018}.

The goal of this Letter is twofold. First, we propose a versatile system, i.e., an array of optical waveguides embedded in an atomic gas, that allows one to combine different physical phenomena in a manageable manner, including topological phase transitions, spontaneous $\PT$-symmetry breaking, and nonlinearity in a nonlinear quadrimer system featuring odd-$\PT$ symmetry. Second, we introduce a nonlinear non-Hermitian quadrimer generalization of the well-known Rice-Mele model~\cite{RiceMele}, that describes the above array, report families of nonlinear modes bifurcating from linear topological edge states, and study stability of nonlinear edge states. We show that each of the degenerated linear mode bifurcates in two distinct nonlinear families, and each nonlinear mode represents a doublet of odd-$\PT$-conjugate states.
The modes in the doublet share the  same propagation constant and power but have different field polarizations. The doublet families generated at the left and right edges are characterized by different existence ranges and stability properties. The findings reported here bring insights to the interplay of non-Hermiticity, topology, and nonlinearity for realizing different phase transitions and nonlinear modes in a universal platform and achieving their active manipulation, promising for applications in optical information processing and transmission.

\begin{figure}[htbp]
\centering
\includegraphics[width=\linewidth]{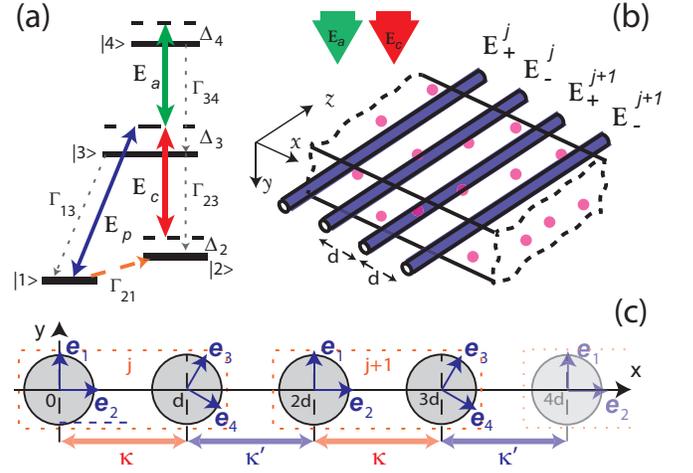}\\
\caption{(a)~The energy-level diagram and the excitation scheme of the inverted-Y type system. The probe ($\mathbf{E}_{p}$), control ($\mathbf{E}_{c}$), and assistant ($\mathbf{E}_{a}$) laser fields drive transitions $|1\rangle\leftrightarrow|3\rangle$,  $|2\rangle\leftrightarrow|3\rangle$, and $|3\rangle\leftrightarrow|4\rangle$, respectively. $\Gamma_{21}$ denotes the incoherent pumping rate from $|1\rangle$ to $|2\rangle$ (providing gain); other $\Gamma_{jl}$ are spontaneous-emission decay rates from $|j\rangle$ to $|l\rangle$. (b)~A possible setup for realizing the target probe-field susceptibility consisting of an array of optical waveguides, equally separated by distance $d$, embedded in the atomic gas. (c)~Polarizations and primitive cells $j$ and $j+1$ (shown by dotted lines) of an array with intra- and inter-cell coupling coefficients $\kappa$ and $\kappa'$.}
\label{fig1}
\end{figure}

\vspace{2mm}
{\it The physical model.--}  A dielectric permittivity of an atomic gas can be modified with great flexibility allowing creation of a prescribed symmetry~\cite{HHK2013,Sheng2013,Peng2016,Zhang2016}. Bearing this in mind, we consider an array of anisotropic optical waveguides, having equal radii $r_w$, embedded in a cold four-level atomic gas with an inverted-Y type configuration [Fig.~\ref{fig1}(a)]. The use of the inverted-Y configuration is to take the advantage of electromagnetically induced transparency~\cite{FIM2005}, which can largely suppress the large absorption of the probe field due to the spontaneous emission of the atoms in the intermediated state $|3\rangle$; additionally, it is useful for realizing the odd-$\PT$ symmetry in the system by tuning the control and assistant fields independently.

Neighboring waveguides are separated by a distance $d$ [Fig.~\ref{fig1} (b)] and are arranged to have principal optical axes mutually rotated in the $(x,y)$ plane by an angle $\alpha$ [Fig.~\ref{fig1} (c)]. Principal optical axes of a pair of waveguides in a primitive cell are determined by two pairs of mutually orthogonal unit vectors $\mathbf{e}_{1,2}$ and $\mathbf{e}_{3,4}$, corresponding to the left and right waveguides in a cell.
The waveguides have equal $x$- and $y$-components of the dielectric tensor but different $z$-components, originating a mismatch $2\delta$ between the propagation constants of the left (``$+$'') and right (``$-$'') waveguides, i.e., $\beta_{\pm}=\beta\pm\delta$ ($\beta$ is average propagation constant).

We apply a control (``$c$'') and assistant (``$a$'') fields $E_{s}=\mathcal{E}_{s}e^{\ii k_{s}y-\ii\omega_{s} t}+{\rm c.c.}$ (hereafter $s=a,c$) to the atomic cell [Fig.~\ref{fig1} (b)]. A transversely polarized probe laser field 
is applied to the waveguides, generating guided modes with mutually orthogonal linear polarizations, holding also outside the waveguides due to weak guidance. Components of the probe field in the $j$th primitive cell can be expressed as
\begin{equation}
	\label{fieldsE}
	\mathbf{E}_{+,-}^{j}=
	\left(\mathbf{e}_{1,3}A_{1,3}^{j}\psi_{1,3}^{j}+\mathbf{e}_{2,4}A_{2,4}^{j}\psi_{2,4}^{j}\right)
	e^{\ii\beta_\pm z-\ii\omega_pt},
\end{equation}
where $A_{1,\ldots,4}^{j}(z)$ are slowly varying amplitudes, and $\psi_{1,\ldots,4}^{j}(\mathbf{r})$ are  normalized transverse distributions.


Under electric-dipole and rotating-wave approximations, the system Hamiltonian in the interaction picture reads $\hat{H}_{\rm int}=\hbar\sum_{j=2}^4[\Delta_j|j \rangle\langle j|-\hbar(\Omega_p|3\rangle\langle
1|+\Omega_c|3 \rangle\langle 2| +\Omega_a|4 \rangle\langle 3|+{\rm h.c.})]$.  Here $\Delta_{j}$ are detunings,  $\Omega_{p}=(\mathbf{e}_z \cdot \mathbf{p}_{31})\mathcal{E}_p/\hbar$, $\Omega_{c}=(\mathbf{e}_z \cdot \mathbf{p}_{32})\mathcal{E}_c/\hbar$, and
$\Omega_{a}=(\mathbf{e}_z \cdot \mathbf{p}_{43})\mathcal{E}_a/\hbar$ are respectively Rabi frequencies of
the probe, control, and assistant fields, with
$\mathbf{p}_{jl}$ being the electric-dipole matrix elements associated with  the transition between that atomic states $|j\rangle$ and $|l\rangle$. The dephasing due to the interaction of atoms near waveguide surfaces may be avoided by using coating paraffin or siloxane on the surfaces~\cite{coating}, or by using the technique of nanofiber-based optical dipole trap~\cite{nanofiber}.
The probe-field susceptibility $\chi_p$ can be obtained by solving the Maxwell-Bloch equations governing the evolution of the atoms and the light field (see~\cite{Suplemental} for details).

Our target is to acquire an $x$-dependent probe-field susceptibility, $\chi_p(x)$, which under the change of the parameters leads either to a hermitian or to a $\PT$-symmetric array. In particular, one  obtains~\cite{Suplemental}:
\begin{equation}
\chi_p =
\chi_w+\ii\chi_\ii
-e^{\ii\phi_{1}}\chi_{1}\cos(2\pi x/d)-e^{\ii\phi_{2}}\chi_{2}\sin(\pi x/d),
\label{target}
\end{equation}
where $\chi_w$ is a real transverse susceptibility of a wave-guide, $\chi_\ii$ describes uniform gain or absorption,  $\chi_{1,2}$  ($|\chi_{1,2}|\ll |\chi_{w}|$) and $\phi_{1,2}$ are respectively amplitudes and phases of the susceptibility modulations.
The distribution~(\ref{target}) can be achieved, e.g., in a gas of laser-cooled $^{87}$Rb atoms
with the levels assigned as $|1\rangle=|5 S_{1/2},\,F=1\rangle$, $|2\rangle=|5 S_{1/2},\,F=2\rangle$, $|3\rangle=|5 P_{3/2},\,F=3\rangle$, $|4\rangle=|6 S_{1/2}\rangle$,
with the atomic parameters given by
$\mathcal{N}_a=3.0\times10^{14}$~cm$^{-3}$ (atomic density), $\Gamma_{13}\approx\Gamma_{23}\approx10^3\Gamma_{34}\approx2\pi\times3$~MHz, $\Gamma_{21}\approx2\pi\times 50$~kHz, $\Delta_2=0.1$~MHz, $\Delta_3=-41$~MHz, and $\Delta_4=0.2$~GHz.

To obtain susceptibility~(\ref{target}) in this system the Rabi frequencies must satisfy~\cite{Suplemental}
\begin{eqnarray}
\label{fields}
{\Omega_{s}}
	\approx {\Omega_{s 0}}
	+\Omega_{s 1}\chi_1\cos(\phi_1+\phi_{s}) \cos(2\pi x/d)
	\nonumber \\
	+\Omega_{s 1}\chi_2
	\cos(\phi_2+\phi_{s})\sin(\pi x/d),
\end{eqnarray}
where $\Omega_{c0}=10\,$MHz, $\Omega_{a0}=20\,$MHz, $\Omega_{c 1}\approx 8\,$MHz, $\Omega_{a 1}\approx 610\,$MHz, $\phi_c\approx 1.1$, and $\phi_a\approx-0.3$
($\chi_{1,2}$ and $\phi_{1,2}$ remain to be free parameters).
The validity of susceptibility (\ref{target})
should be limited to a certain domain, say
for an array of 20 waveguides with $d=4r_w=4\,\mu$m.

\vspace{2mm}
{\it A non-Hermitian nonlinear quadrimer lattice.--} For cylindrical waveguides in the tight-binding approximation one has $\psi_{1,2}^{(j)}({\bm r})\approx\psi(\bfr)$ and $\psi_{3,4}^{(j)}({\bm r} )
\approx \psi(x-d,y)$.
The probe field in the array [Fig.~\ref{fig1} (b,c)] is governed by the equation for the amplitude column-vector $\mathbf{A}^{j}=(A_{1}^{j}, A_{2}^{j}, A_{3}^{j}, A_{4}^{j})^{\rm T}$ (${\rm T}$ means  transposition):
\begin{equation}
\label{chain}
\ii\frac{\mathrm{d}{ \bA}^{j}}{\mathrm{d}z}=
H\bA^{j}+\kappa'(H_-\bA^{j-1}+H_+\bA^{j+1})
-F(\bA^{j})\bA^{j}.
\end{equation}
Here $H=\delta\sigma_3\otimes\sigma_0+\kappa (H_++ H_-)$, $\sigma_{1,2,3}$ are the Pauli matrices, $\sigma_0$ is the $2\times 2$ identity matrix, $H_-=H_+^{\rm T} = \frac{\ii}{2}(\sigma_1 + \ii\sigma_2)\otimes \rR_\alpha$, and $\rR_\alpha$ ia the matrix of 2D rotation by the angle $\alpha$. The intra-cell ($\kappa$) and inter-cell ($\kappa'$) coupling coefficients are given by
~\cite{Suplemental}:
\begin{eqnarray}
	\label{coupling}
	\{\kappa\,,\kappa'\}
	=
	\frac{k_p}{2 \ii}\int_{jd}^{(j+1)d}\!\! \mathrm{d}x\int_{-\infty}^\infty \! \! \mathrm{d} y\, \psi(\sqrt{x^2+y^2})
	\nonumber\\
	\quad\times[\chi_w-\chi_p(x-jd)]
	\psi(\sqrt{(x-d)^2+y^2}),
\end{eqnarray}
where
$j=0$ ($j=1$) stands for $\kappa$ ($\kappa'$).
The diagonal  focusing Kerr  nonlinearity matrix reads
\begin{eqnarray}
	\label{nonlinearity}
	F(\bA^{j})={\rm diag}\left(|A_{1}^{j}|^2+\frac{2}{3}|A_{2}^{j}|^2,|A_{2}^{j}|^2
	+\frac{2}{3}|A_{1}^{j}|^2, \right.
	\nonumber\\
	\left. |A_{3}^{j}|^2+\frac{2}{3}|A_{4}^{j}|^2,|A_{4}^{j}|^2+\frac{2}{3}|A_{3}^{j}|^2\right).
\end{eqnarray}

By changing the parameters $\chi_{1,2}$ and $\phi_{1,2}$ one can obtain different symmetries and phases of chain (\ref{chain}). In particular, if $(\phi_1,\phi_2)=(\pi/2,\pm\pi/2)$ and $\chi_{\ii}=0$ in (\ref{target}),
$\kappa$ and $\kappa'$ are both real and the array is non-Hermitian; we call it odd-$\PT$-chain. Such chain features odd-$\p\T_f$-symmetry~\cite{Jones-Smith2010,Bender2011,Konotop2018} with the parity operator $\p=\sigma_3\otimes\sigma_0$ and  odd-time reversal (or fermionic) operator $\T_f=\ii\sigma_0\otimes\sigma_2\K$,  $\K$ being complex conjugation ($\T_f^2=-1$). If $\phi_1=0$ and $\phi_2=0,\pi$, 
then $\kappa$ and $\kappa'$ are both imaginary, and the array is Hermitian; we call it h-chain. An h-chain is $\T_f$- and  $\T$-symmetric with $\T=\K$ ($\T^2=1$). Furthermore, 
one can obtain topologically trivial ($|\kappa|>|\kappa'|$) and topologically nontrivial ($|\kappa|<|\kappa'|$) phases by choosing, respectively, $\phi_2=\pi/2$ ($\phi_2=\pi$) and $\phi_2=-\pi/2$ ($\phi_2=0$) for an odd-$\PT$-chain (h-chain). Without loss of generality, we assume $\delta \geq 0$, and  $\kappa, \kappa'\geq 0$ (for odd-$\PT$-chain) and Im$\,\kappa$, Im$\,\kappa'\geq 0$ (for h-chain).

The relation between the parameters of the gas and different topological phases follows from the fact that the linear Hamiltonian of (\ref{chain}) [at $F(\bA^j)=0$] in the momentum space ($\bA^j=\ba_q e^{\ii qj}$), $H(q)$, can be block-diagonalized by the unitary transformation
	\begin{equation*}
		\label{block}
		UH(q)U^\dag=\sigma_0\otimes h(q),
\quad 		
U=\left(
		\begin{array}{cccc}
			-\sin\alpha & \cos\alpha & 0 & 0
			\\
			0 & 0 & 0 & 1
			\\
			\cos\alpha & \sin\alpha & 0 & 0
			\\
			0 & 0 & 1 & 0
		\end{array}\right),
	\end{equation*}
where $h(q)=\ii(\kappa+\kappa'\cos q)\sigma_1+\ii\kappa'\sin (q)\sigma_2+\delta \sigma_3.$
In the Hermitian case,   $h(q)$ is the celebrated Rice-Mele Hamiltonian~\cite{RiceMele,RevModNiu} whose
non-Hermitian generalization is also known~\cite{WaZSo}. Thus, Eq.~(\ref{chain}) can be viewed as a {\it  nonlinear non-Hermitian quadrimer generalization of the Rice-Mele model}, where $A_{1,...,4}^j$ can be treated as ``internal'' degrees of freedom of $j$th primitive cell. Topological properties of the linear limit of  (\ref{chain}) are determined by the topological properties of $h(q)$. Indeed, let $\ba_q$ and $\tba_q$ be the eigenvectors of $H(q)$ and $H^\dag(q)$ constituting a biorthonormal basis ($\tba_{q^\prime}^\dag\ba_q=\delta_{q^\prime q}$;  hereafter  $\delta_{mj}$ is the Kronecker symbol), and the Zak phase for a given band is defined by $\varphi=\int_{BZ}\tba_q^\dagger\partial_q\ba_q \mathrm{d}q$, where the integral is over the Brillouin zone. Let also $\bal_q$ and $\tbal_q$ be the eigenvectors of $h(q)$ and $h^\dag(q)$ constituting a biorthonormal basis ($\tbal_{q^\prime}^\dag\bal_q=\delta_{q^\prime q}$), while $\varphi_h=\int_{BZ}\tbal_q^\dagger\partial_q\bal_q \mathrm{d} q$ is the Zak phase of the Rice-Mele lattice defined by $h(q)$. Then the block-diagonal structure of $UH(q)U^\dag$ implies that $\ba_q= (1,1)^{\rm T}\otimes \bal_q/\sqrt{2}$ and, respectively, $\varphi=\varphi_h$. The phases $\varphi_h$ are computed explicitly in~\cite{Suplemental}.


Eigenvalues of the linear limits of both h-chain and odd-$\PT$-chain are doubly degenerate:
there are two branches of bulk modes  $\pm \tb(q)$, where
$
\tb(q)=[\delta^2-\kappa^2-\kappa^{\prime 2}-2\kappa\kappa'\cos q]^{1/2}.
$
An odd-$\PT$-chain can belong to unbroken ($\delta>\delta_1\equiv\kappa+\kappa'$),  {\it partially} broken ($\delta_2\equiv|\kappa-\kappa'|<\delta<\delta_1$), or  {\it fully} broken ($\delta<\delta_2$)  phase. Since $\tb(q)=\tb(-q)$, in the case at hand the skin-effect~\cite{Skin1,Skin2} is prevented by the symmetry~\cite{Kawabata19} (which does not exclude that effect subject to different properties of the quadrimers obtained by proper configurations of the external fields).

\vspace{2mm}
{\it Nonlinear edge states.--}  Semiinfinite chains are obtained by truncation of (\ref{chain}).  For the left-edge (``L'') chain we consider  $\bA_L^{0,1,\ldots}$ 
assuming  $\bA_L^{-1}=0$. For the right-edge (``R'') chain we consider $\bA_R^{0,-1,\ldots}$
assuming $\bA_R^{1}=0$. In the linear limit of the topologically nontrivial phase, $|\kappa/\kappa'|<1$,  at each edge there exist two independent edge states:   
$\bA_{L,m}^{j}=\left(-{\kappa}/{\kappa'}\right)^je^{-\ii\delta z }(\delta_{m1},\delta_{m2},0,0)^{\rm T}
$ and $\bA_{R,m}^{j}=\left(-{\kappa'}/{\kappa}\right)^{j}e^{\ii \delta z}
(0,0,\delta_{m1},\delta_{m2})^{\rm T}$, with
$m=1,\,2$.

Nontrivial topology by itself is not necessary for existence of nonlinear edge states~\cite{BluKon,Hoq}. Meantime,  nonlinear modes can bifurcate from the linear topological edge states $\bA_{L,m}^{j}$ and $\bA_{R,m}^{j}$, when topological characteristics of the bifurcating families are uniquely associated with the topological numbers of the underling linear lattice. Such modes will be called {\it nonlinear topological edge states}.
Below we focus on the families of solutions in the odd-$\PT$-chain.
Owing to the odd-$\PT$-symmetry-protected degeneracy of guided modes, in the small-amplitude limit, nonlinear edge states can be searched as a superposition:
$\bA_{g}^j \approx\epsilon e^{\ii\epsilon^2\lambda z}(\bA_{g,1}^j \sin\nu + \bA_{g,2}^j \cos\nu)$,
where $g=L$ (L-modes) or $g=R$ (R-modes), $\epsilon\ll1$ is a formal small parameter, $\lambda$ is the nonlinearity-induced shift of the propagation constant, and $\nu$ is a parameter to be determined. The   perturbation analysis~\cite{ZezKon13,Konotop2018,Suplemental,BF} reveals two cases when the bifurcations are allowed.
In the first case $\nu=\pi/4$, i.e., the bifurcation occurs from the linear superposition of two states $\bA_{g,1}$ and $\bA_{g,2}$  with equal ``weights'', and $\lambda = 5\kappa'^2/[6(\kappa'^2+\kappa^2)]$. In the second case $\nu=0$, i.e., the nonlinear mode bifurcates from only one linear edge state and $\lambda =  \kappa'^2/(\kappa'^2+\kappa^2)$.

The above small-amplitude edge states serve as an initial guess for the numerical investigation of entire families of stationary nonlinear edge states whose dependence on $z$ is  $\bA_g^{j}(z) \propto e^{\ii b z}$, where $b$ is the real nonlinear propagation constant. Distinct families are characterized by the dependencies of the total dimensionless power $P_g=\sum
(\bA_g^j)^\dagger\bA_g^j$ on $b$.

\begin{figure}[htbp]
	\centering
	\includegraphics[width=\linewidth]{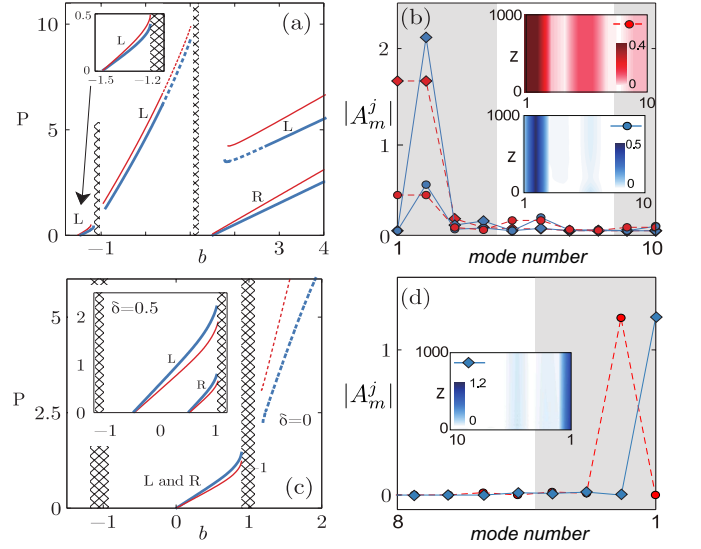}
	\caption{(a) Families of nonlinear edge states in the odd-$\p\T$-chain with unbroken $\p\T$ symmetry for $(\kappa,\kappa')=(0.1,1)$ and $\delta=1.5$. Hatched areas indicate the spectral bands. The blue thick and red thin lines correspond to $\nu=0$ and $\nu=\pi/4$, respectively. Solid and dashed curve fragments correspond to stable and unstable edge states. The inset zooms in families bifurcating from the left linear edge states. (b) Examples of L-modes in the lower semi-infinite gap with $b=-1.25$ (blue solid and red dashed lines with circles) and in the upper semi-infinite gap with $b=3$ (blue solid and red dashed lines with diamonds); the lower and upper insets show respectively stable propagation of L-modes in blue solid and red dashed lines with circles. (c) Families of nonlinear edge states in the h-chain for $(\kappa,\kappa')= (0.1\ii,\ii)$ and $\delta=0$ (main panel) or $\delta=0.5$ (inset). (d) Example of R-mode in the odd-$\p\T$-chain with $b=3$ (blue solid line with diamonds), and its $\p\T_f$-partner (red dashed line with circles); the inset illustrates stable propagation of the R-mode in blue solid line with diamonds. In all panels $\alpha=\pi/6$.}
	\label{fig2}
\end{figure}

In the unbroken $\PT$-symmetric phase ($\delta>\delta_1$), the linear spectrum of (\ref{chain}) consists of two bands separated by a central finite gap [Fig.~\ref{fig2}(a)]. In topologically nontrivial phase there exist two linear edge states in the semi-infinite gaps. The families of L-  and R-modes bifurcate from the linear edge states in the lower and upper semi-infinite gaps, respectively. There also exist families of nonlinear edge states having nonzero excitation power threshold, as illustrated in Fig.~\ref{fig2}(a).
L-modes exist only in a relatively narrow interval of $b$ between $-\delta$ and the lower edge of the first band. These modes become delocalized when they approach the first band and ``reappear'' in the finite gap. Meanwhile, nonzero-threshold families of L-modes exist also in the upper semi-infinite gap, where two zero-threshold families of R-modes emerge in the upper semi-infinite gap. Using linear stability analysis and direct propagation method, we find that all nonlinear modes shown in Fig.~\ref{fig2}(a) are stable except for small segments plotted with dashed lines.  In contrast to the more common even $\PT$ symmetry with $\T^2=1$, the odd time reversal implies that any nonlinear edge state represents a {\it doublet}, i.e., the pair $(\bA, \p\T_f \bA)$,
characterized by different polarizations of the fields $\bE_\pm^j$ in Eq.~(\ref{fieldsE}).  Remarkably, due to the symmetry both   modes in each doublet  are stable (or unstable) simultaneously.

For comparison, in Fig.~\ref{fig2}(c) we show families of nonlinear edge states in an h-chain, where bifurcation occurs from  edge states in the finite gap. For $\delta=0$ (the quadrimer SSH-limit) we observe families of doublets bifurcating from the zero-energy topological linear edge state. The degeneracy is lifted at $\delta\neq 0$ [inset in  Fig.~\ref{fig2}(c),  where L- and R-modes have different power dependencies].

\begin{figure}[tbp]
\centering
\includegraphics[width=\linewidth]{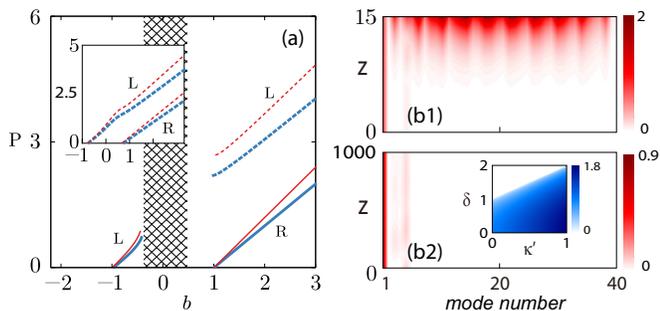}
\caption{(a) Families of nonlinear edge states in the partially broken odd-$\p\T$ phase with $\delta=1$; the inset shows the families in the fully broken $\p\T$ phase with $\delta=0.8$. Hatched area corresponds to the linear band. The blue thick and red thin lines correspond to $\nu=0$ and $\nu=\pi/4$, respectively. (b1) Unstable propagation of the L-mode in the partially broken $\p\T$ phase.
(b2) Quasi-stable propagation of the same mode as in (b1) but in the presence of additional absorption with $\gamma=0.5$; inset shows domains of unbroken (white) and broken (blue) $\p\T$ phases in the $(\kappa',\delta)$-panel with $\kappa=1$. The colorbar shows the maximum of $\text{Im}[\tb_+(q)]$. Other parameters are the same as in Fig.~\ref{fig2}(a).}
\label{fig3}
\end{figure}

In Fig.~\ref{fig3}(a) we show families bifurcating from linear topological edge states in the broken $\PT$-symmetric phase. These families behave similarly to those in the unbroken case, except that now there exists only one linear band due to the closing of the gap. 
In the fully broken $\PT$-phase ($\delta<\delta_2$) the families of the L-modes are continuous [inset in Fig. 3(a)] since the linear band consists of  purely imaginary propagation constants.

Unlike in the case of unbroken $\PT$-symmetric phase where the instability is triggered only by nonlinearity, in broken $\PT$-symmetric phases all nonlinear edge states are \textit{apriori} unstable, because a continuum of unstable linear waves inevitably destabilizes the ``tails'' of edge states, as  exemplified in Fig.~\ref{fig3}(b1). Remarkably, even in a fully broken $\PT$-phase, nonlinear edge states can be almost stabilized by introducing additional absorption in the central part of the lattice. The introduction of such absorption (which is localized in the region of exponentially decaying tails of the edge states) aims to suppress  the growth of unstable bulk modes but has no significant affect on the edge states themselves. This determines the strength of the absorption $\gamma$ which must be equal to the largest increment of the bulk modes, characterized by the maximum of $\text{Im}[\tb_+(q)]$, i.e., $\gamma=\text{max}(\text{Im}[\tb_+(q)])$.
By introducing such absorption as an additional term $-i\gamma\bA^{j}$ on the right-hand side of Eq.~(\ref{chain}), the L-mode can propagate robustly up to $z=1000$, as illustrated in Fig.~\ref{fig3}(b2).

\vspace{2mm}
{\it Conclusion.--} We have introduced a nonlinear non-Hermitian quadrimer generalization of the Rice-Mele model, with a primitive cell characterized by four degrees of freedom. The model describes an array of anisotropic waveguides embedded in a gas of cold atoms, and represents a universal platform for implementation of several types of phase transitions in a single setting, as well as different symmetries in the nonlinear regime. Varying external laser fields enables transitions between Hermitian and non-Hermitian configurations, as well as between trivial and nontrivial topological phases, both in linear and nonlinear regimes. We have considered the cases when the chain is either Hermitian or non-Hermitian featuring odd-time-reversal symmetry, the latter supporting doublets of nonlinear states. The modes in the doublet are characterized by different light polarizations but identical stability properties. Families of nonlinear topological modes are unambiguously related to the topological linear edge states they are bifurcating from. Even in the broken $\PT$-symmetric phase, the observation of nonlinear edge doublets is possible by using additional stabilizing absorption in the central part of the array.

\vspace{2mm}
\textit{Acknowledgments}. C. H. was supported by the by the National Natural Science
Foundation of China (No. 11974117), the National Key Research and Development
Program of China (Nos. 2016YFA0302103 and 2017YFA0304201), and Shanghai Municipal Science and
Technology Major Project (No. 2019SHZDZX01). D. A. Z. was supported by the Foundation for the Advancement of Theoretical Physics and Mathematics ``BASIS'' (No. 19-1-3-41-1). G. H. acknowledges financial support from the National Natural Science
Foundation of China (No. 11975098). V. V. K. acknowledges financial support from the Portuguese Foundation for Science and Technology (No. UIDB/00618/2020).
	


\end{document}